\documentclass[10pt,twocolumn,letterpaper]{article}

\usepackage{cvpr}
\usepackage{times}
\usepackage{epsfig}
\usepackage{graphicx}
\usepackage{amsmath}
\usepackage{amssymb}
\usepackage{enumitem}

\usepackage[pagebackref=true,breaklinks=true,letterpaper=true,colorlinks,bookmarks=false]{hyperref}

\cvprfinalcopy 

\def\cvprPaperID{****} 

\ifcvprfinal\fi
\begin{document}

\title{A Systematic Study of Deep Learning Models and xAI Methods for Region-of-Interest Detection in MRI Scans}

\author{
\small Justin Yiu$^{1,2}$, 
Kushank Arora$^{1}$, 
Daniel Steinberg$^{1}$, 
Rohit Ghiya$^{1}$ \\
\\
\small$^{1}$Georgia Institute of Technology, USA \\
\small$^{2}$Imperial College London, United Kingdom \\
\\
{\tt\small wyiu31@gatech.edu, karora72@gatech.edu, dsteinberg8@gatech.edu, rghiya7@gatech.edu}
}

\maketitle

\begin{abstract}
   Magnetic Resonance Imaging (MRI) is an essential diagnostic tool for assessing knee injuries. However, manual interpretation of MRI slices remains time-consuming and prone to inter-observer variability. This study presents a systematic evaluation of various deep learning architectures combined with explainable AI (xAI) techniques for automated region of interest (ROI) detection in knee MRI scans. We investigate both supervised and self-supervised approaches, including ResNet50, InceptionV3, Vision Transformers (ViT), and multiple U-Net variants augmented with multi-layer perceptron (MLP) classifiers. To enhance interpretability and clinical relevance, we integrate xAI methods such as Grad-CAM and Saliency Maps. Model performance is assessed using AUC for classification and PSNR/SSIM for reconstruction quality, along with qualitative ROI visualizations. Our results demonstrate that ResNet50 consistently excels in classification and ROI identification, outperforming transformer-based models under the constraints of the MRNet dataset. While hybrid U-Net + MLP approaches show potential for leveraging spatial features in reconstruction and interpretability, their classification performance remains lower. Grad-CAM consistently provided the most clinically meaningful explanations across architectures. Overall, CNN-based transfer learning emerges as the most effective approach for this dataset, while future work with larger-scale pretraining may better unlock the potential of transformer models.\\

\end{abstract}

\section{Introduction}

\subsection{\textbf{Background}}MRI is a cornerstone of modern medical diagnostics, widely used for non-invasive and high-resolution visualization of soft tissues. However, interpreting MRI scans—especially reviewing each individual slice to identify abnormalities—can be a labor-intensive and time-consuming process for medical professionals \cite{Edelstein10}. For instance, in diagnosing a meniscus injury, clinicians must manually review an entire scan sequence, isolate the critical slices, and focus on the region of interest (ROI) to make accurate decisions \cite{Sladky11}. This manual diagnostic process is not only inefficient but also prone to variability. Automating ROI detection in MRI scans could significantly improve diagnostic speed and reliability. In particular, knee MRI scans are frequently used to diagnose a range of conditions such as meniscus tears, ACL injuries, and joint degeneration. In our research, we will be focusing on the meniscus—cartilage structures that serve as shock absorbers between the femur and tibia (see Figure 1) \cite{LondonBridge23}.

\begin{figure}[h]
\centering
\includegraphics[width=0.8\linewidth]{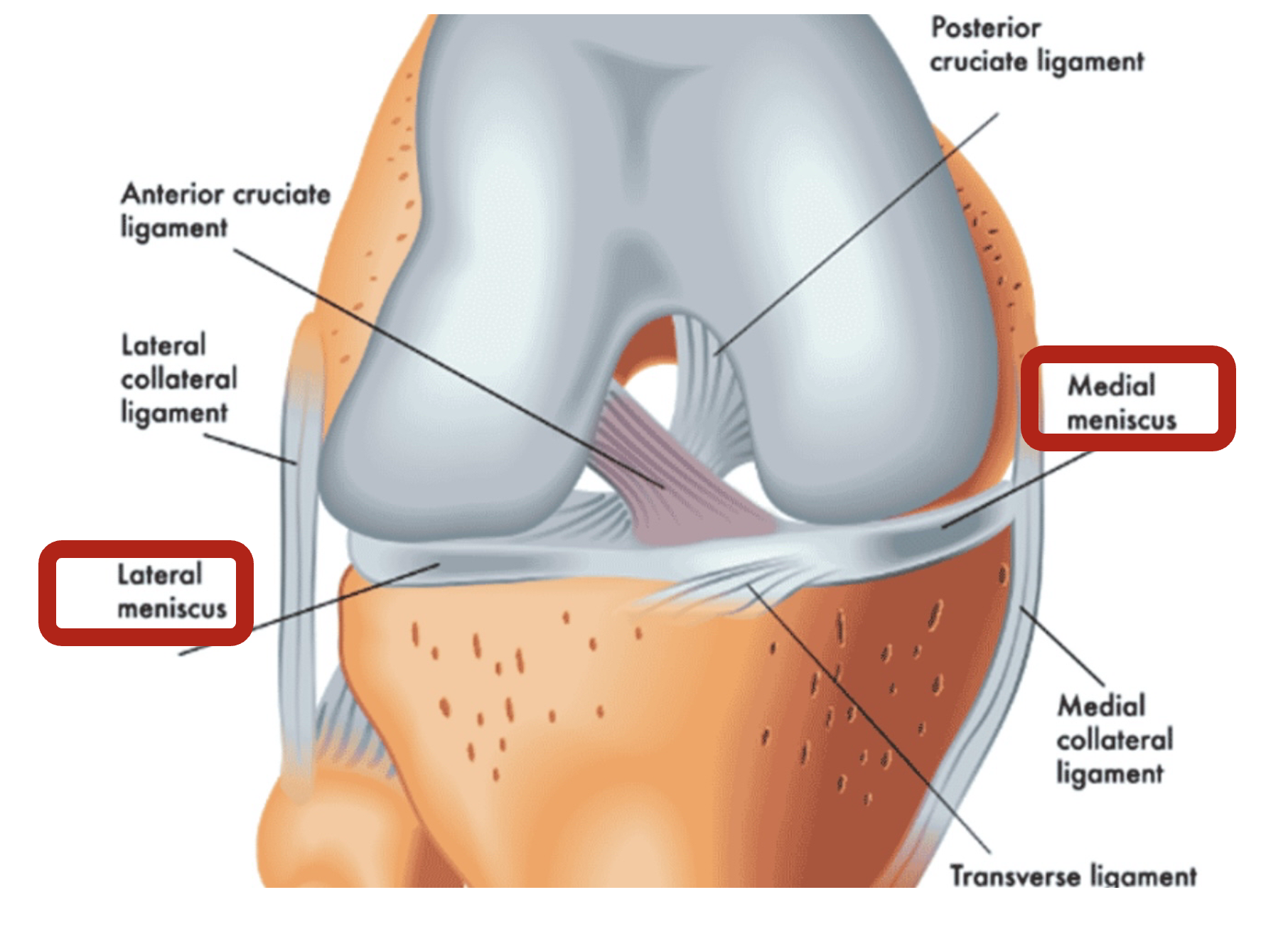}  
\caption{Image visualizing meniscus of a knee.}
\label{fig:kneeMRI}
\end{figure}

\subsection{\textbf{Objectives}} This study aims to develop and compare multiple deep learning model architectures for effective ROI detection in knee MRI scans, with a specific focus on meniscus injuries. In addition to evaluating model performance, we integrate xAI techniques—such as Grad-CAM and Saliency Maps—to enhance model interpretability and ensure that decision-making processes are transparent and clinically meaningful. Both quantitative metrics and qualitative assessments (e.g.,  accuracy of ROI highlighting in scans) will be used to evaluate model performance. Ultimately, we target to identify the most effective combinations of model architectures and xAI methods that can reliably support diagnostic decision-making in a clinical setting.

\subsection{\textbf{Related Works}} Deep learning methods—particularly convolutional neural networks (CNNs)—have been widely applied in medical imaging. U-Net architectures are commonly used for segmentation tasks, while CNNs such as ResNet have shown strong performance in classification tasks when trained on annotated datasets \cite{Pedoia21}. Vision Transformers (ViTs) have also gained attention due to their ability to model global contextual relationships, especially when pretrained on large-scale medical or general-purpose datasets. In recent years, researchers have increasingly explored the integration of explainability techniques—such as Grad-CAM and Saliency Maps—to visualize the regions influencing model predictions \cite{Qian24} \cite{Zhang19}. These methods are aiming to help both clinicians and researchers better understand the rationale behind a model’s output and open up the possibility of accelerating diagnosis, particularly if the model can accurately highlight the region of interest (ROI) \cite{Qian24}.

\subsection{\textbf{Current Limitations}} Despite recent advancements, several critical limitations persist in the application of deep learning to knee MRI analysis. First, many models suffer from shallow interpretability; without the use of xAI techniques, it remains unclear which image features the models rely on for diagnosis. Second, most existing approaches prioritize classification accuracy over precise localization, resulting in weak ROI detection and often requiring manual preprocessing steps during inference \cite{Pedoia21}. Third, the availability of annotated datasets—particularly for rare tear types—is limited, which restricts model generalizability. Fourth, current models are sensitive to domain shifts, such as variations in MRI scanners or imaging protocols, leading to potential performance degradation in real-world applications \cite{Zhang19}. Finally, while some studies, such as Lasagni et al. \cite{Lasagni25} and Musthafa et al. \cite{Musthafa24}, have combined ResNet with Grad-CAM, and others like Mulat et al. \cite{Mulat24} have adopted U-Net with Saliency Maps, there is no unified or standardized framework for evaluating and comparing these different model-xAI combinations using both clinical and technical criteria.

\subsection{\textbf{Motivation}} To address the limitations of existing work, we propose a systematic evaluation of deep learning architectures and xAI methods for knee MRI analysis. This study aims not only to benchmark model performance but also to assess clinical interpretability, thereby bridging the gap between algorithmic accuracy and diagnostic utility. This research is of particular interest to several key stakeholders in the medical and AI communities. Clinicians and radiologists stand to benefit from AI-assisted tools that can highlight diagnostically relevant regions in MRI scans, reducing diagnostic workload, minimizing human error, and improving diagnostic throughput. Healthcare providers and hospitals may also gain operational efficiency, especially in high-demand settings where expert radiological review is limited. Furthermore, AI researchers and developers in the medical imaging domain will benefit from this study’s insights into which architectures and xAI techniques provide meaningful, interpretable outputs in real-world diagnostic contexts. If successful, our research could contribute to the development of AI models that not only achieve high diagnosis accuracy but also provide transparent and clinically relevant visual justifications for their predictions. In the longer term, such models may accelerate early diagnosis, reduce unnecessary follow-up imaging, and contribute to more personalized and data-driven treatment planning in orthopedics medicine.

\subsection{\textbf{Data}} 
We used the knee MRI dataset provided by Stanford University Medical Center - MRNet~\cite{origpaper}, designed to support the development of automated diagnostic systems for prioritizing high-risk patients and assisting clinicians. The data were collected between January 1, 2001, and December 31, 2012, using various MRI machines and protocols, resulting in scans with 17 to 61 slices (mean: 31.48; SD: 7.97). The dataset comprises MRI data from 1,199 patients (1,088 in training and 111 in validation), with clinical indications including acute or chronic pain, injury, trauma, and pre- or post-operative evaluation. The average patient age is 38 years, and 41.5\% are female. All scans are grayscale images stored in \texttt{.npy} format, which facilitates integration with Python-based workflows via NumPy. Access to the dataset requires acceptance of a Research Use Agreement~\cite{StanfordMLGroup} prohibiting commercial use, redistribution, and re-identification. The dataset is not FDA-approved and is not intended for clinical applications.

The dataset is self-contained and widely used in medical imaging research. It includes 1,130 training and 120 validation scans, with a hidden test set reserved for benchmarking against the original baseline. To prevent data leakage, all scans from a given patient are assigned to a single split. Each split contains at least 50 positive cases per label. Validation labels were determined by majority vote from three musculoskeletal radiologists. The training set is organized by anatomical plane (axial, coronal, sagittal), while the validation set includes binary labels for abnormalities, ACL tears, and meniscal tears. In the dataset, 80.6\% of exams are abnormal; 23.3\% show ACL tears, 37.1\% show meniscal tears, and 38.2\% exhibit both.

\section{Approach}

While previous studies have developed deep neural networks for ROI detection in medical imaging, there is a lack of systematic comparisons—particularly involving xAI techniques. In this study, we evaluate several custom-designed deep learning architectures to detect ROI in knee MRI scans, focusing specifically on meniscus injuries. To ensure consistency and relevance, we restricted our analysis to the sagittal plane, which provides a clear cross-sectional view of the meniscus. A key contribution of our work is the novel integration of a self-supervised U-Net with a multi-layer perceptron (MLP) classifier, combined with xAI techniques such as Grad-CAM for interpretability—an approach not systematically studied in prior research. We expected this to be effective, as the U-Net encoder captures rich spatial features during reconstruction, which the classifier can then leverage for diagnosis. xAI methods further validate the clinical relevance of predictions, supporting both accuracy and transparency. Overall, our work fills an important research gap by offering a systematic comparison of model types and providing actionable insights for future efforts to automate ROI detection in MRI scans.\\

The MRNet dataset, sourced from Stanford University Medical Center \cite{MRNet}, includes 1,130 training and 120 validation knee MRI samples, with 35\% of the training set (397 cases) showing meniscal tears—indicating a slight class imbalance. Each input is a grayscale image of shape (s, 256, 256), where s denotes the number of slices per patient and varies across samples. We used this dataset to explore both supervised and self-supervised approaches for detecting ROIs within each MRI slice. The overall model design, illustrated below, generalizes across all experiments.

\begin{figure}[!htb]
\centering
    \includegraphics[width=0.48\textwidth,]{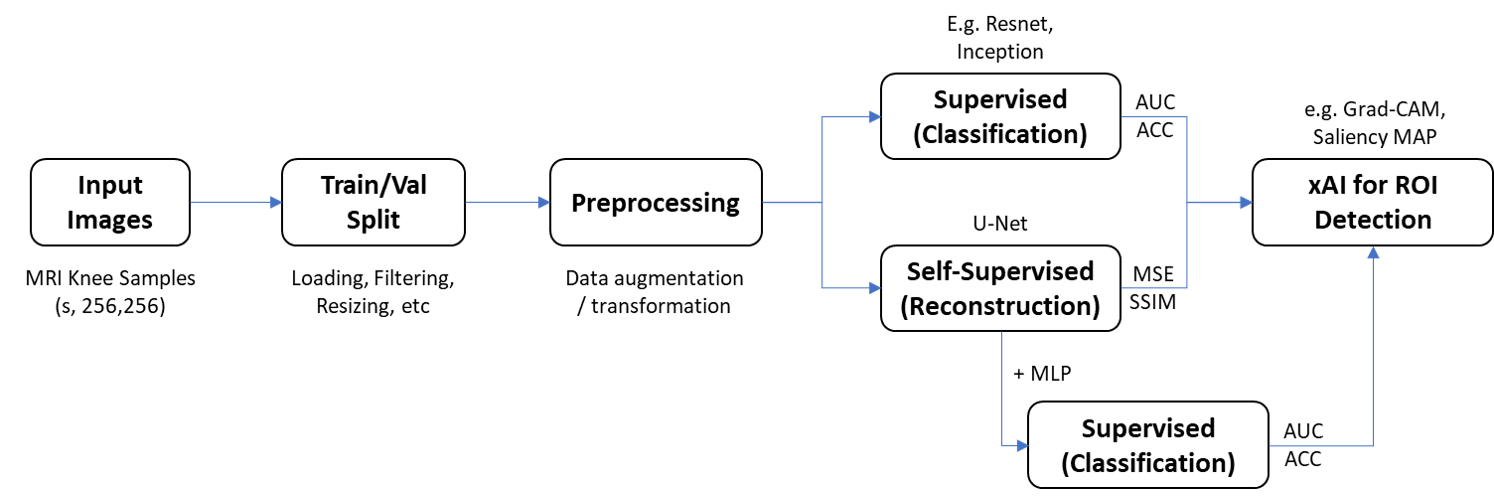}
\caption{\label {} Model design}
\end{figure}

As shown in Figure 2, we employed multiple methods to detect ROI for meniscus tears within each MRI slice for every patient.

\subsection{\textbf{Supervised : Classification}}

Area under the curve (AUC) of ROC was chosen as the validation metric due to its robustness with imbalanced data and to align with the original study for comparison. As the input has varying number of slices, a batch size of 1 is used for ResNet50 and InceptionV3 to squeeze the input to use each slice as an input to the CNN with the number of slices being used as the number of samples per batch for the pretrained CNN. 

\subsubsection{\textbf{ResNet50}}
To improve upon the original AlexNet-based approach, ResNet50 was selected for its strong performance and computational efficiency. To address overfitting, a dropout layer was added to the pretrained model. Data augmentation included random rotations (±25°), shifts (±25 pixels), and horizontal flipping (p=0.5), consistent with the original setup \cite{origpaper}. Padding and interpolation methods for standardizing MRI slice counts were also tested but added computational overhead without performance gains. Nine models were trained, one for each label-anatomical plane pair. Two inference strategies were evaluated: a single-model prediction and majority voting across planes; the latter performed slightly better overall. 

\subsubsection{\textbf{InceptionV3}}
The model uses a customized pre-trained InceptionV3 \cite{InceptionV3}, with added dropout layers between convolutional blocks for improved regularization. A global average pooling layer feeds into a sigmoid-activated linear layer for final output. Data augmentation follows the original paper \cite{origpaper}, and hyperparameters were manually tuned. The best model had AUC lower than the original AlexNet implementation. Overfitting persisted despite regularization \& tuned hyperparameters, likely due to InceptionV3's depth and the limited training data. 

\subsubsection{\textbf{Vision Transformer (ViT)}}
This is a two-stage transformer model designed to classify sequences of knee MRI slices. An Image Encoder, based on a pretrained ViT\cite{rw2019timm, steiner2021augreg, dosovitskiy2020vit}, processes each MRI slice individually to extract its feature embedding. A Sequence Encoder, which is a standard Transformer, then takes the sequence of these image embeddings as input. A special [CLS] token is added to the sequence to aggregate information across all slices. The final output of this [CLS] token is passed through a linear layer to produce the classification (meniscus tear or not). The model is also built to extract attention maps from both encoders for visualization and interpretability.

\subsection{\textbf{Self-Supervised : U-Net}}
This model explores the feasibility of learning generalizable, clinically relevant features from knee MRI scans without labels. Using a U-Net architecture for image-to-image reconstruction, we apply xAI techniques to visualize encoder-extracted features in the latent space. Reconstruction quality is assessed using peak signal-to-noise ratio (PSNR) and structural similarity index measure (SSIM) to evaluate whether the model captures critical structural information.

{\renewcommand{\arraystretch}{1.1} 
\begin{table*}[t!]
\begin{center}
\begin{tabular}{|l|l|l|l|}
\hline
 \multicolumn{2}{|c|}{\textbf{Architecture: ResNet50}} & 
 \multicolumn{2}{|c|}{\textbf{Architecture: InceptionV3 Variants}} \\
\hline
\textbf{Hyperparameter} & \textbf{Search Space} & \textbf{Hyperparameter} & \textbf{Search Space} \\
\hline
Learning Rate & {1e-2, 1e-4, 1e-5} & Learning Rate & {1e-1, 1e-3, 1e-4} \\
\hline
Dropout Ratio & {0.5 - 0.75} & Regularization Coeff. & {1e-1, 1e-3, 5e-4} \\
\hline  Epoch & {10}  & Epoch & {20, 50}  \\
\hline

 \multicolumn{2}{|c|}{\textbf{Architecture: Vision Transformer (ViT)}} & 
 \multicolumn{2}{|c|}{\textbf{Architecture: U-Net Variants}} \\

\hline
\textbf{Hyperparameter} & \textbf{Search Space} & \textbf{Hyperparameter} & \textbf{Search Space} \\
\hline
Learning Rate & {1e-5, 1e-6} & Upsampling Method &
Bilinear Interpolation, \\ 
& & & Transposed Convolution \\

\hline
Transformer Depth & {4, 8} & Activation Function & ReLU, LeakyReLU  \\
\hline
Transformer Heads & {8, 12} & Base channels: & {32, 64, 128} vs {64, 128, 256}  \\ 
\hline
\end{tabular}
\end{center}
\caption{Hyperparameter Search Space for Model Optimization}
\label{tab:hyperparameter_tuning}
\end{table*}

\subsection{\textbf{U-Net (Self-Supervised) + MLP Classifier}}

This model family combines U-Net’s spatial encoding-decoding capabilities with a MLP classifier to jointly perform reconstruction and binary classification. The encoder’s output is shared between the MLP (for diagnosis) and the decoder (for image reconstruction). Grad-CAM is applied to the classification head for ROI visualization. Input slices are normalized to $[0, 1]$, resized to $224 \times 224$, and trained using a combined loss function: binary cross-entropy (BCE) for classification and MSE for reconstruction. AUC is the primary classification metric, while PSNR and SSIM assess reconstruction performance.

\subsubsection{U-Net + MLP Classifier}

This variant uses a custom U-Net with an encoder based on pretrained ResNet-18 or ResNet-101~\cite{ResNet18}, paired with a two-layer MLP classifier attached to the latent representation. This setup supports dual-task learning but increases computational complexity and requires regularization to prevent overfitting.

\subsubsection{Pretrained U-Net + Residual MLP Classifier}

This model uses a pretrained U-Net released by Meta~\cite{fastmri}, with a residual MLP classifier attached to the frozen or fine-tuned encoder. The encoder includes convolutional blocks with instance normalization, dropout, and LeakyReLU activation, followed by average pooling. The decoder uses transposed convolutions with skip connections and $1 \times 1$ convolutions. This design benefits from faster convergence and better generalization, though domain adaptation to MRNet-specific features may be limited due to fixed pretraining.

\subsection{\textbf{xAI for ROI Detection}}

We experimented with various explainable AI (xAI) techniques—Grad-CAM, Saliency Maps, Guided Grad-CAM, SmoothGrad, and Guided Backpropagation—to visualize spatial regions influencing model predictions through heatmaps, particularly for detecting meniscus tears. For pretrained classification models (Section 2.1), heatmaps were generated from the global average pooling layer following the last convolutional block and overlaid on the original 256×256 image. For self-supervised models with a classification head (Section 2.3), xAI techniques were applied to the classification output. In purely reconstruction-based models, the encoder’s final convolutional layer (prior to pooling) was used for heatmap generation. Among all methods, Grad-CAM produced the most coherent and interpretable results and was therefore chosen as the primary technique for qualitative comparison across models.

\subsection{\textbf{Anticipated and Encountered Problems}}
We anticipated several key challenges in this work. First, we expected high computational costs and long training times, as prior studies have reported that training deep neural networks on MRI data can take several days \cite{origpaper}. Second, we anticipated overfitting due to the limited availability of labeled MRI data and the high capacity of modern CNN architectures. Both of these challenges were indeed encountered during our experiments. Training was time-consuming even with access to top-tier hardware (NVIDIA H100 GPUs), and early models showed signs of overfitting, particularly on the more complex architectures. To mitigate these issues, we employed pretrained models to leverage transfer learning and applied extensive data augmentation to increase training data variability. None of our initial models worked out of the box. However, as we iteratively refined the architectures and incorporated the above countermeasures, they began to yield promising results—details of which are presented in the following sections.

\section{Experiments and Results}
This section presents the comprehensive experimental framework, evaluation metrics, and results for both the self-supervised reconstruction and supervised classification tasks. We provide a detailed analysis of the quantitative and qualitative outcomes, justify the methodological decisions made during experimentation, and discuss the performance of various models and techniques employed.

\subsection{Experimental Setup}
To ensure reproducibility and rigorous evaluation, we established a standardized experimental setup. All models were trained and evaluated on the MRNet dataset's images for sagittal view, which was already partitioned into training and validation sets. The primary objective of our experimentation was to perform a systematic hyperparameter search to identify the optimal configuration for each model architecture.

\subsubsection{Hyperparameter Tuning}
We conducted a comprehensive grid search over a predefined hyperparameter space for each model. The choice of hyperparameters was informed by a combination of literature review, architectural best practices, and empirical exploration. Hyperparameter space for some of the models is shown in Table \ref{tab:hyperparameter_tuning}.

\subsection{Evaluation Metrics}
Model success was evaluated using quantitative metrics and qualitative analysis.

\begin{itemize}
    \item \textbf{Supervised Classification Metrics} Due to class imbalance in knee injury classification, \textbf{Area Under the Receiver Operating Characteristic Curve (AUC)} was the primary metric, robust to imbalance ($1.0$ perfect, $0.5$ random). \textbf{Accuracy} is also reported.
    \item \textbf{Self-Supervised Reconstruction Metrics} For image reconstruction, the goal is fidelity to the original. We used:
    \begin{itemize}
        \item \textbf{Peak Signal-to-Noise Ratio (PSNR)} which measures signal power to noise ratio; higher is better.
        \item \textbf{Structural Similarity Index Measure (SSIM)} which measures perceptual similarity considering luminance, contrast, and structure ($-1$ to $1$, with $1$ being perfect).
    \end{itemize}
\end{itemize}

\subsection{Explainable AI (xAI) for Region of Interest (ROI) Identification}
A key objective was to move beyond classification to interpretation by identifying the ROIs that informed the model's decision. We evaluated five gradient-based attribution methods to generate saliency maps.

\subsubsection{Comparison of xAI Techniques}

\begin{figure}[!htb]
\centering
\includegraphics[width=0.48\textwidth,]{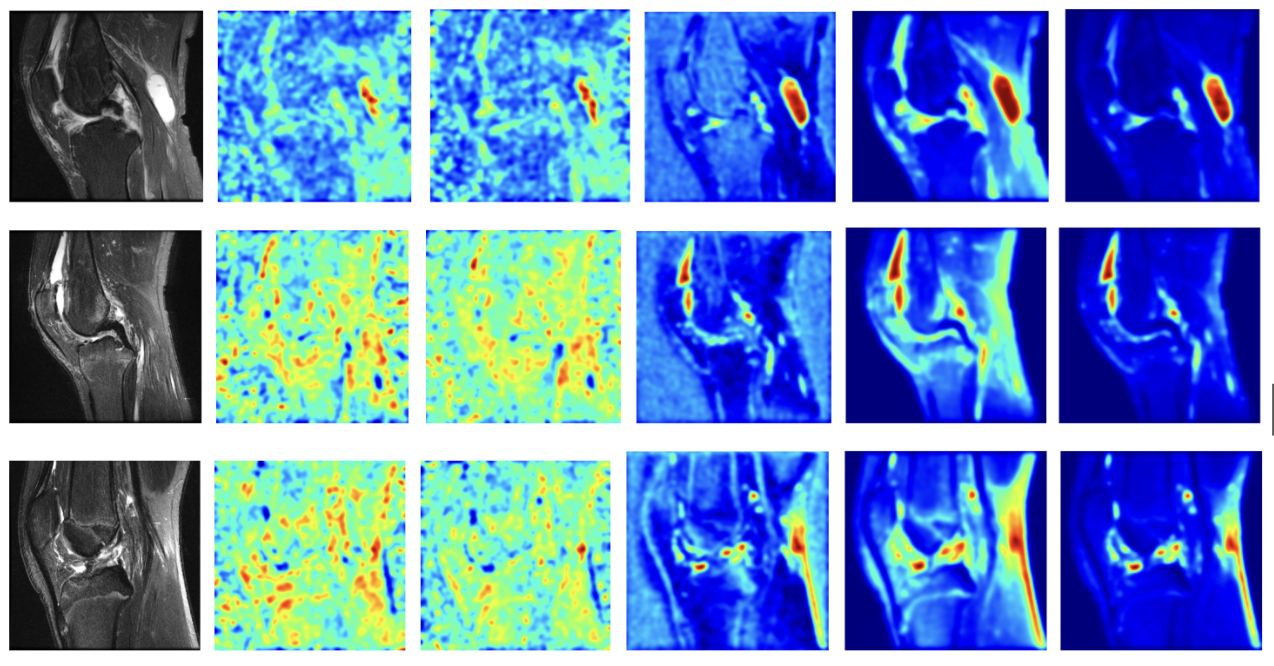}
\caption{\label {fig:roi_comp} Comparison of xAI Techniques (Original MRI Scan, Saliency Map, SmoothGrad, GuidedBackProp, GradCAM, GuidedGradCAM from left to right) from three different Input Scans}
\end{figure}

The following techniques were implemented and compared on our best-performing classification model: \textbf{Saliency Maps}, which visualizes the gradient of the output category with respect to the input pixels; \textbf{SmoothGrad}, which averages the gradients over multiple noisy versions of the input image to produce a cleaner saliency map; \textbf{Guided Backpropagation}, which combines standard backpropagation with a ReLU layer modification to only backpropagate positive gradients; \textbf{Grad-CAM}, which uses the gradients flowing into the final convolutional layer to produce a coarse localization map highlighting important regions; and \textbf{Guided Grad-CAM}, an element-wise product of Guided Backpropagation and Grad-CAM, aiming for high-resolution, class-discriminative visualizations.

\begin{figure*}[t!]
\centering
    \includegraphics[height=0.32\textwidth]{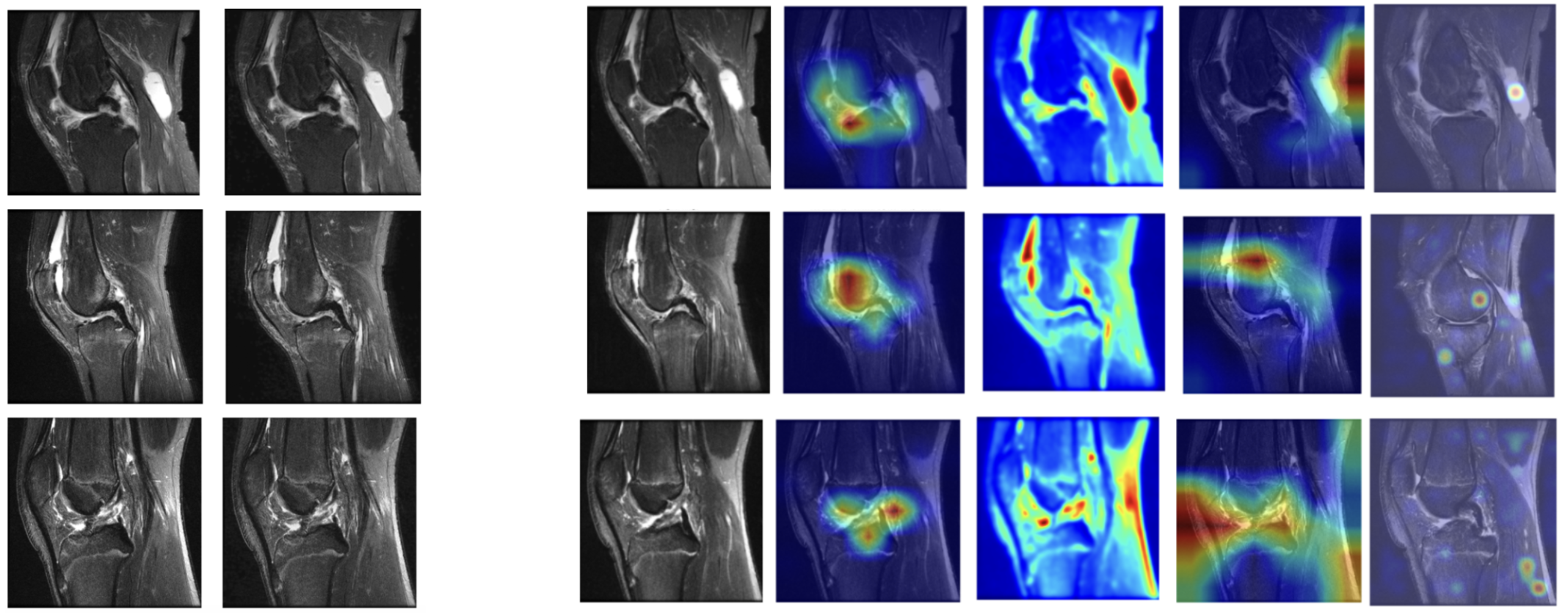}
\caption{\label {fig:roi_models_comp} Qualitative comparison: \textbf{Left}: Original MRI Scan (column 1) and their corresponding reconstructions (column 2). \\ \textbf{Right}: Comparative Grad-CAM maps from Original MRI Scan, ResNet50, U-Net, InceptionV3, and ViT, in order}
\end{figure*}

\begin{figure*}[!htb]
\centering
\includegraphics[height=0.23\textwidth,]{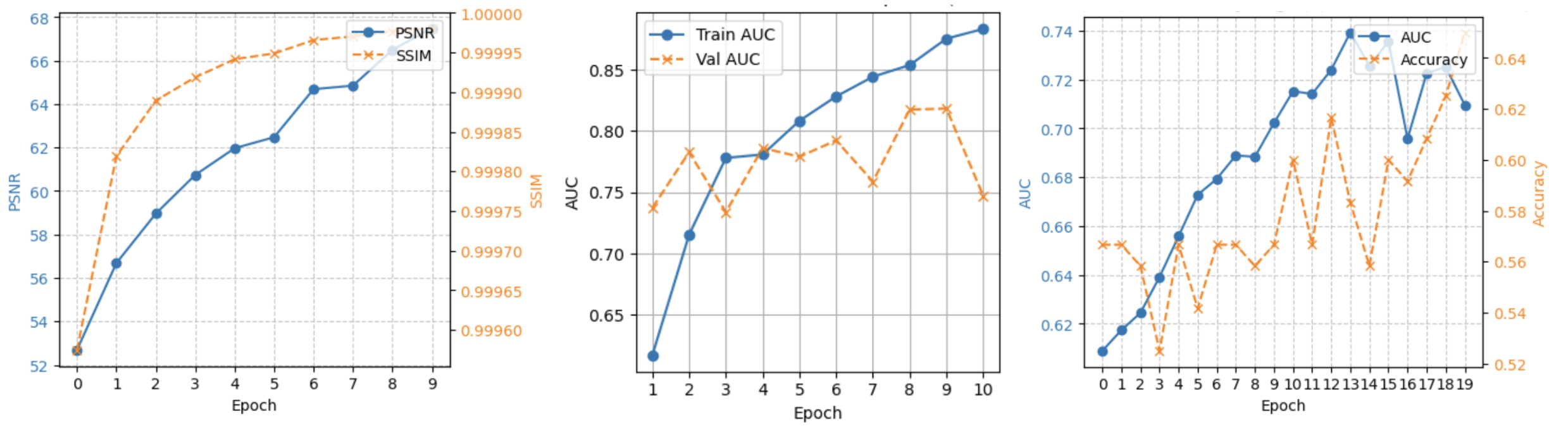}
\caption{\label {fig:curves} From Left to Right: (i) PSNR and SSIM for Self-Supervised U-Net (ii) AUC for Train vs Validation for Resnet50 (iii) AUC and Accuracy curves for ViT-based classifier }
\end{figure*}

\subsubsection{Qualitative Results and Selection}
Based on Figure \ref{fig:roi_comp} Saliency Maps produced noisy and difficult-to-interpret visualizations, while Grad-CAM provided clear, localized heatmaps (albeit coarse), Guided Grad-CAM offered the most visually precise and well-defined heatmaps for explaining model decisions. However, due to the complex integration of Guided Grad-CAM's custom activations into pre-trained architectures, which risked altering learned representations, \textbf{Grad-CAM} was ultimately chosen for the final analysis across all models. This decision prioritised a robust and reliable method for identifying regions of interest without requiring significant architectural modifications or risking unintended changes to the models.

\subsection{Experimental Results}

\begin{table*}[t]
\centering
\begin{tabular}{|l|c|c|c|c|}
\hline
\textbf{Model} & \textbf{AUC} & \textbf{Accuracy} & \textbf{PSNR} & \textbf{SSIM} \\
\hline
ResNet50 (supervised)    & \textbf{0.8184} & 0.74 & --   & -- \\
InceptionV3 (supervised) & 0.72            & 0.66 & --   & -- \\
Vision Transformer (ViT) & 0.74            & 0.67 & --   & -- \\
U-Net (self-supervised)  & --              & --   & 67.5 & 0.99998 \\
U-Net + MLP (hybrid)     & 0.725           & 0.65 & 34.0 & 0.97 \\
\hline
\end{tabular}
\vspace{6pt} 
\caption{Results Comparison Across All Models}
\label{tab:consolidated_results}
\end{table*}

\subsubsection{Supervised Classification Results}
We evaluated three primary classification models—ResNet50, InceptionV3, and Vision Transformer (ViT)—after fine-tuning on the MRNet dataset. ResNet50 achieved the strongest performance, with an AUC of 0.8184 and an accuracy of 0.74, outperforming both InceptionV3 (AUC = 0.72, Accuracy = 0.66) and ViT (AUC = 0.74, Accuracy = 0.67). ResNet’s residual connections and pretrained initialization allowed it to effectively capture deep hierarchical features while generalizing well to limited medical data. By contrast, InceptionV3 and ViT underperformed, likely due to their higher capacity and greater reliance on larger datasets or domain-specific pretraining to realize their full potential.

\subsubsection{Self-Supervised (with MLP Classifier)}

\textbf{U-Net - Reconstruction Results: }For the self-supervised setting, a custom U-Net trained for image reconstruction achieved strong fidelity, with PSNR = 67.5 and SSIM = 0.99998, indicating robust structural feature learning. However, the highest-PSNR variant showed signs of overfitting, with unstable validation curves. The most stable configuration employed transposed convolution for upsampling, a learning rate of 1e-4, and CNN channels of [64, 128, 256].\\

\noindent\textbf{U-Net + MLP Classifier Results: }When extended with an attached MLP classifier, the U-Net encoder produced an AUC of 0.725 and an accuracy of 0.65. As expected, reconstruction quality dropped (PSNR = 34, SSIM = 0.97) since the model balanced dual objectives of reconstruction and classification. While this hybrid approach did not match ResNet50’s classification accuracy, it demonstrated that spatially rich self-supervised features can be repurposed for diagnostic prediction.\\

Qualitative comparisons in Figure \ref{fig:roi_models_comp} reinforced these trends: ResNet50 consistently highlighted clinically relevant regions, while ViT yielded less reliable localization maps.

\subsection{Bias-Variance Analysis}
An important aspect of model evaluation is understanding the bias–variance trade-off:

\begin{itemize}
    \item \textbf{ResNet50} demonstrated low variance, consistently achieving the highest AUC across multiple runs. Its inductive bias toward local spatial features allowed it to generalize well from the limited MRNet dataset. 
    \item \textbf{InceptionV3} showed signs of high variance, with performance fluctuating despite extensive regularization. This suggests it may require larger datasets to stabilize training.
    \item \textbf{Vision Transformer (ViT)} exhibited relatively high bias, likely due to the model’s reliance on large-scale pretraining and substantial data, which were not fully available in this study. The underperformance suggests that ViTs underfit when constrained to small datasets like MRNet.
    \item \textbf{U-Net + MLP hybrid} models balanced bias and variance depending on the task: strong reconstruction metrics but weaker classification performance indicate that the encoder learned structural features well but did not optimize fully for the classification task.
\end{itemize}

This analysis provides theoretical context for why CNNs, particularly ResNet50, outperformed ViTs in our experiments. It also clarifies why the self-supervised + MLP models are best seen as exploratory baselines rather than definitive diagnostic solutions.

\subsection{Conclusion}
In conclusion, our experiments successfully identified high-performing models for both MRI reconstruction and classification. Transfer learning and strong encoder priors improved classification, while reconstruction benefited primarily from architecture and training stability. Among classification models, ResNet50 consistently outperformed other architectures for meniscus tear diagnosis and ROI detection. InceptionV3 and Vision Transformers underperformed relative to ResNet, underscoring the difficulty of training deeper or transformer-based models effectively on limited medical datasets. While transformer-based models theoretically offer advantages through their global receptive fields, in our setting these benefits did not materialize, likely due to insufficient dataset size and lack of domain-specific pretraining.\\

The U-Net + MLP hybrid models demonstrated promise in leveraging spatial features for reconstruction and interpretability, but their classification accuracy lagged behind CNNs. These results suggest that while hybrid and transformer approaches remain interesting directions, CNN-based transfer learning remains the most practical and effective method for knee MRI classification under current data constraints.\\

Finally, although our study provides meaningful comparisons, we recognize its limitations: the number of models and configurations tested was modest. Future work should include a more exhaustive set of experiments, particularly with larger-scale pretrained ViTs, expanded hyperparameter searches, and bias–variance characterization across different dataset splits. Such extensions would help determine whether transformer-based models could eventually surpass CNN-based methods in this domain.

\section{Work Contribution}

All authors contributed equally to this research.


{\small
\bibliographystyle{ieee_fullname}
\bibliography{egbib}
}

\end{document}